# Fundamental Theories and Epistemic Shifts: Can History of Science Serve as a Guide?

Helge Kragh[*]

**Abstract**: The present discussion concerning certain fundamental physical theories (such as string theory and multiverse cosmology) has reopened the demarcation problem between science and non-science. While parts of the physics community see the situation as a beginning epistemic shift in what defines science, others deny that the traditional criterion of empirical testability can or should be changed. As demonstrated by the history of physics, it is not the first time that drastic revisions of theory assessment have been proposed. Although historical reflection has little to offer modern physicists in a technical sense, it does offer a broader and more nuanced perspective on the present debate. This paper suggests that history of science is of some indirect value to modern physicists and philosophers dealing with string theory, multiverse scenarios, and related theoretical ideas.

The recent debate about fundamental physical theories with no or little connection to experiment and observation concerns primarily the relationship between theoretical physics and philosophy. There are reasons to believe that a more enlightened perspective on the debate can be obtained by taking into regard also the history of physics and history of science generally. Possibly unknown to many physicists, there are several historical precedents, cases which are somewhat analogous to the present one and from which much can be learned. Apart from outlining what I consider to be the essence of the current debate, this paper briefly discusses the general role that history of science can play in science and philosophy of science. It refers to some noteworthy lessons from past physics, of which one particular case, namely the nineteenth-century vortex theory of matter, is singled out as a possible analogue to

---

the methodological situation in string physics. While I do not suggest that these earlier cases are substantially similar to the ones concerning string theory and the multiverse, I do suggest that there are sufficient similarities on the level of methodology and rhetoric to make them relevant for modern physicists and philosophers.

**1. The very meaning of science**

In May 2008 there appeared in *New Scientist* an article with the provocative question "Do we need to change the definition of science?" Six years later *Nature* included an appeal to "Defend the integrity of physics" (Matthews, 2008; Ellis & Silk, 2014). Both articles discussed essentially the same question, namely whether or not certain recent developments in theoretical physics belong to science proper. For more than a decade there has been an ongoing and often heated dispute in the physics community, and also in some corners of the philosophical community, concerning the scientific status of theories such as superstring physics and multiverse hypotheses. These theories are cultivated by a fairly large number of professional physicists and are thus, by ordinary sociological standards, undoubtedly to be counted as scientific. But are they also scientific from an epistemic point of view or does their status as branches of physics require an extension or revision of the traditional meaning of science?

The classical demarcation problem between science and non-science (which may or may not include pseudoscience) has taken a new turn with the appearance of fundamental and highly mathematical theories which may not be experimentally testable in the ordinary sense. So why believe in them? According to the philosopher Dudley Shapere (2000, pp. 159-61), "physics is in fact approaching, or perhaps has reached, the stage where we can proceed without the need to subject our further theories to empirical test." He asks, "Could empirical enquiry, which has guided up to a certain point science in its history, lead at that point to a new stage wherein empiricism itself is transcended, outgrown, at least in a particular domain?" There are more than a few physicists who would presently respond affirmatively to Shapere's question. It should be noted that the demarcation problem and the traditional criteria of falsifiability and empirical testability are not only discussed by physicists but also in some other branches of science. For example, biologists have questioned these criteria and suggested, in striking analogy to the debate concerning multiverse physics, that methodological norms of what constitutes good science are not only irrelevant but also detrimental to the progress of their science (Yang, 2008).

What it is all about can be summarized in the notion of "epistemic shifts," meaning claims that the basic methodological and epistemological rules of science are in need of revision (Kragh, 2011). These rules may be appropriate for most science and have been appropriate for all science until recently; but in some areas of modern physics they are no longer adequate and should therefore be replaced by other norms for the evaluation of theories. A proposed shift in epistemic standards may be of such a drastic nature that it challenges the very meaning of science as traditionally understood. In this case it effectively implies a new demarcation line separating what counts as science and what does not. This is what Steven Weinberg (2007) alluded to when he, referring to the string-based multiverse, said that "we may be at a new turning point, a radical change in what we accept as a legitimate foundation for a physical theory."

Another way of illustrating the notion of an epistemic shift is to compare it to Thomas Kuhn's idea of revolutions separated by different paradigms. Richard Dawid (2013, p. 124) speaks of the debate in the physics community as "a paradigm shift regarding the understanding of scientific theory assessment." According to the original version of Kuhn's philosophy of science paradigm shifts include different criteria for what counts as acceptable science and also for evaluating theories. Rival paradigms carry with them rival conceptions of science and for this reason alone they are incommensurable. In principle, no rational argument can decide whether one paradigm is superior to a competing paradigm. The rhetoric of epistemic shifts has become part of modern physics. "We are in the middle of a remarkable paradigm shift in particle physics," asserts one physicist, referring to the anthropic string landscape (Schellekens, 2008). And according to another physicist, the multiverse promises "a deep change of paradigm that revolutionizes our understanding of nature" (Barrau, 2007).

The purpose of this paper is not to re-examine the recent debate concerning string theory and multiverse cosmology but rather to look at it through the sharp lenses of history of science. Although knowledge of the history of the physical sciences is of no direct relevance to the ongoing debate, it is of some indirect relevance. It may serve the purpose of correcting various mistakes and to place the subject in a broader historical perspective. Physicists may think that superstrings and the multiverse have ushered in a uniquely new situation in the history of science, but if so they are mistaken. There have been several cases in the past of a somewhat similar nature if not of quite the same scale. I modestly suggest that modern fundamental physics can in some sense learn from its past. Before turning to this past I shall briefly review what is generally and for good reasons considered the most important of the traditional standards of theory evaluation, namely that a theory must be testable.

## 2. Testability

To speak of the "definition" of science is problematic. There simply is no trustworthy methodological formulation which encapsulates in a few sentences the essence of science and is valid across all periods and all disciplines. Nonetheless, there are some criteria of science and theory choice which are relatively stable, enjoy general acceptance, and have been agreed upon since the early days of the scientific revolution (Kuhn, 1977). Almost all scientists subscribe to the belief that testability is more than just a desideratum that scientists have happened to agree upon and which suited science at a certain stage of development. They consider it a *sine qua non* for a theory being scientific that it must be possible to derive from it certain consequences that can be proved right or false by means of observation or experiment. If there are no such consequences, the theory does not belong to the domain of science. In other words, although empirical testability is not a sufficient criterion for a theory being scientific, it is a necessary one.

Einstein was a great believer in rationalism and mathematical simplicity and yet he was convinced that "Experience alone can decide on truth" (Einstein, 1950, p. 17). He is followed by the large majority of modern physicists who often go to great lengths in order to argue that their theories, however speculative and mathematical they may appear to be, do connect with empirical reality. Lee Smolin (2004, p. 70) echoed Einstein when he concluded about the opposing views of string theory and loop quantum gravity that, "Because this is science, in the end experiment will decide."

Physicists working with string theory, multiverse cosmology or related areas of fundamental physics are routinely accused of disregarding empirical testability and to replace this criterion with mathematical arguments. The accusations are not quite fair (Johansson & Matsubaru, 2009; Dawid, 2013, p. 154). By far most physicists in these fields readily accept the importance of testability, admitting that empirical means of assessment have a higher epistemic status than non-empirical means. On the other hand, they stress the value of the latter methods which sometimes may be the only ones available. At the same time they maintain that their theories have – or in the near future will have – consequences that at least indirectly can be tested experimentally. They have not really abandoned the commonly accepted view of experiment as the final arbiter of physical theory. "The acid test of a theory comes when it is confronted with experiments," two string theorists say (Burgess & Quevedo, 2007, p. 33). Unfortunately the necessary experiments are in most cases unrealistic for the time being, but what matters to them is that predictions from the theories are not beyond empirical testability in principle.

Although one can identify a consensus view concerning testability, it is to some extent rhetorical and of limited practical consequence. One thing is to agree that theories of physics must be testable, but another thing is the meaning of the concept of testability, where there is no corresponding consensus. Everyone agree that actual and present testability, involving present instrument technologies or those of a foreseeable future, is preferable, but that is where the agreement ends. Some of the questions that physicists and philosophers have discussed are the following.

(1) Should it be required that a theory is actually testable, or will testability in principle, for example in the form of a thought experiment, suffice?
(2) Should a theory result in precise and directly testable predictions, or will indirect testability do?
(3) If a theory does not result in precise predictions, but only in probability distributions, is it then testable in the proper sense?
(4) Will a real test have to be empirical, by comparing consequences of the theory with experiments or observations, or do mathematical consistency checks also count as tests?
(5) Another kind of non-empirical testing is by way of thought experiments or arguments of the *reductio ad absurdum* type. A theoretical model may lead to consequences that are either contradictory or so bizarre that they are judged unacceptable. How should such arguments enter the overall evaluation picture?
(6) At what time in the development of a theory or research programme can one reasonably demand testability? Even if a theory is not testable at present, in a future version it may result in testable predictions.
(7) How should testability be weighed in relation to other epistemic desiderata? For example, is an easily testable theory with a poor explanatory record to be preferred over a non-testable theory with great explanatory power? What if the testable theory is overly complicated and the non-testable one is mathematically unique and a paragon of simplicity?
(8) Should predictions of novel phenomena be counted as more important than pre- or retrodictions of already known phenomena?

I shall not comment further on these issues except pointing out that some of them are particularly relevant with regard to theories, such as string theory or multiverse scenarios that are not actually testable in the ordinary way. Indirect testability (#2) may mean that if a fundamental and well-established background theory with great empirical success results in a prediction which cannot be tested directly, then the success of the background theory functions as an indirect test. The existence of multiple universes can in this sense be said to be tested by quantum mechanics as a background theory. "Accepting quantum mechanics to be universally true means that you should also believe in parallel universes," says Max Tegmark (2007, p. 23).

With respect to #4 there are indeed physicists who have appealed to mathematical consistency as a kind of test of string theories. If admitted as a test it is part of what Dawid (2013) calls non-empirical theory assessment. On the other hand, the issues ##6, 7 and 8 are of a general nature and relevant to all scientific theories whether modern or old. They can be easily illustrated by means of concrete cases from the history of science, but this is beyond the scope of the present paper.

## 3. Physics, philosophy, and history

The fundamental question of the demarcation between science and non-science is of a philosophical and not of a scientific nature. It rests on certain standards and criteria that nature herself does not provide us with and therefore cannot be determined by purely scientific means. The standards and criteria do not need to be part of a philosophical system or even to be explicitly formulated, but they nevertheless belong to the realm of philosophical discourse. The borderline between physics and philosophy has shifted over time, typically with physics appropriating areas which were traditionally considered subjects of philosophical speculation. At least, this is how some physicists like to see the historical development. They question if there is any need for external norms of science of a philosophical nature, suggesting that such norms are unnecessary and may even be harmful to the progress of science. Only the scientists themselves can decide what the boundary is between science and non-science.

According to Barrau (2007), "If scientists need to change the borders of their own field of research it would be hard to justify a philosophical prescription preventing them from doing so." Susskind (2006) agrees, adding that "Good scientific methodology is not an abstract set of rules dictated by philosophers." However, the question of what constitutes a legitimate scientific theory cannot be left entirely to the scientists. The seductive claim that science is what scientists do is circular; moreover, it presupposes that all scientists have the same ideas of what constitutes science. But if there were such a methodological consensus in the physics community there would be no controversy concerning the scientific legitimacy of research areas such as superstrings and the multiverse. History of science strongly suggests that certain methodological prescriptions, such as testability of theories, are almost universally accepted; but it also shows that in some cases the consensus breaks down. The current discussion about string theory and the multiverse is evidently such a case.

In so far that philosophical ideas about science are intended to relate to real science (and not to be purely normative) they must agree with or at least confront the large pool of data provided by the history of science. This is particularly important for philosophical views concerning the dynamics of science or the development of science in its temporal dimension. Consider a philosophical rule which has the consequence that the change from the Bohr-Sommerfeld quantum theory to the new quantum mechanics in the mid-1920s was not progressive. Such a rule is simply not credible.

It may be tempting to consider the history of science as providing an empirical data base for the testing of philosophical theories about science, in somewhat the same way that experiments and observations function as tests of scientific theories. However, this is problematic for several reasons (Schickore, 2011). Foremost among these reasons is that historical case studies exhibit a great deal of variation over time, discipline, and culture. They deal with a particular case at a particular time, with the result that the methodological lessons one can draw from two important cases in the history of science are rarely the same. Philosophers, on the other hand, aim at saying something general about science and its development.

There have been several attempts among philosophers of science to extract rules of scientific development from the history of science or to formulate rules strongly guided by history. None of them have been very successful. Larry Laudan (1977, p. 160) suggested that "our intuitions about historical cases can function as decisive touchstones for appraising and evaluating different models of rationality, since we may say that it is a necessary condition of rationality that it squares with (at least some of) our intuitions." Laudan's standard intuitions included that after 1840 it was irrational to believe that light consisted of particles and similarly, after 1920, that the chemical atoms had no parts. Such examples are however rather trivial and of no real use. The interesting cases in history and philosophy of science are precisely those about which there are no standard intuitions of rationality. Was it rational to believe in the late 1820s in the corpuscular theory of light or, in 1900, in the non-composite atom?

A more recent and more ambitious research project on the testing of theory dynamics included sixteen historical case studies ranging from the Copernican revolution in astronomy to the electroweak unification in particle physics (Donovan, Laudan & Laudan, 1988). These cases were compared with philosophical rules such as "the appraisal of a theory depends on its ability to solve problems it was not invented to solve" and "the standards for assessing theories do not generally vary from one paradigm to another." Although the project resulted in interesting analyses, it failed in establishing a non-trivial philosophical methodology on the basis of history of science. Among its obvious weaknesses was that the project rested on a rather arbitrary selection of case studies; had other cases been chosen the result would have been different.

The failure of the mentioned approach to philosophy of science does not imply that philosophers can or should avoid engaging in historical studies. Indeed, if philosophers want to retain contact with real science and how it changes in time they must take history of science seriously; they must investigate real science in either its present or past forms. By far most of our knowledge of how science has developed comes from the past and is solidly documented in the historical sources. History of science is an indispensable but also, by its very nature, an incomplete guide to understanding contemporary science. By investigating cases from the past and relating them to cases from the present, philosophers as well as active scientists may obtain a broader and more enlightened view of current science.

Historical arguments and analogies have a legitimate albeit limited function in the evaluation of current science. Even a physicist with no interest in or knowledge of the history of his or her field of research cannot avoid being guided by the past. In science as in other areas of human activity it would be silly to disregard the historical record when thinking about the present and the future. On the other hand, such guidance should be based on historical insight and not, as is often the case, on arbitrary selections from a folk version of history. Generally speaking, the history of science is so diverse and so complex that it is very difficult to draw from it lessons of operational value for modern science.

In a paper of 1956 the brilliant and controversial astrophysicist Thomas Gold, one of the fathers of the steady state theory of the universe, argued that cosmology was a branch of physics, hence a science. But he dismissed the idea of a methodology particular to physics or to the sciences generally. "In no subject is there a rule, compliance with which will lead to new knowledge or better understanding," Gold (1956) wrote. "Skilful observations, ingenious ideas, cunning tricks, daring suggestions, laborious calculations, all these may be required to advance a subject. Occasionally the conventional approach in a subject has to be studiously followed; on other occasions it has to be ruthlessly disregarded." Undoubtedly with the ongoing controversy between the steady state theory and relativistic evolution cosmology in mind, Gold further reflected on the lessons of history of science with regard to the methods of science. But he considered history to be an unreliable and even treacherous guide:

> Analogies drawn from the history of science are frequently claimed to be a guide [to progress] in science; but as with forecasting the next game of roulette, the existence of the best analogy to the present is no guide whatever to the future. The most valuable lesson to be learned from the history of scientific progress is how misleading and strangling such analogies have been, and how success has come to those who ignored them.

Gold's cynical and anarchistic view is not without merit, but it seriously underestimates the power of history and the value of insight based on the history of science. Although the development of history of science is not governed by law or exhibits much regularity, neither is it an accidental series of events comparable to a game of roulette.

**4. Some lessons from past physics**

As mentioned, empirical testability is an almost universally accepted criterion of science. But even with respect to this most sacred of the defining features of science we find in the history of science a few exceptions. It is after all not a necessary ingredient of science. Dawid (2013, p. 97) argues that the role played by non-empirical theory assessment in modern fundamental physics is a continuation of earlier tendencies to be found in post-World War II particle physics. This is undoubtedly correct – think of the development of S-matrix or "bootstrap" theory in the 1950s and 1960s – but in my view there is no reason to restrict the historical perspective to the era of quantum and relativity physics. There are inspiration and instruction to be found also in other and earlier examples from the history of physics.

During the early decades of the nineteenth century Romantic natural philosophy (known as *Naturphilosophie*) made a great impact on physics and the other sciences in Northern Europe (Cunningham & Jardine, 1990; Kragh, 2011, pp. 26-34). In this grand attempt to revolutionize science and base it on an entirely new foundation, speculations and aesthetic sentiments were not only considered legitimate parts of science; they were necessary parts and even more fundamental than empirical investigations. The philosopher Friedrich Schelling, the spiritual leader of the *Naturphilosophie* movement, even founded a *Journal of Speculative Physics* as a means of promoting and communicating the new science. At the time the term "speculation" did not have the pejorative meaning it typically has today but was largely synonymous with "intuition." It was a fundamental assumption of the new speculative physics that mind and nature co-existed as a unity, such that one was unable to exist without the other.

Schelling and those who followed his thinking were not necessarily against experiments, but they thought that measuring the properties of objects and phenomena was of no great importance since it provided no understanding of the inner working of nature. In some cases natural philosophers went so far as to completely deny that observation and experiment could lead to any real insights into nature's secrets. The sort of nature that could be empirically investigated was regarded as a dull wrapping that contained and obscured the real, non-objective nature. The only way to recognize the latter was by taking the route of speculative physics, namely, guided by the intuitive mind of the genius. The laws of nature were thought to coincide with the laws of reason; they were true a priori and for this reason it made no sense to test them by means of experiment.

Before dismissing Romantic natural philosophy as nothing but pseudoscientific and metaphysical nonsense it should be recalled that some of the greatest physicists of the time were much influenced by the movement. Examples include H. C. Ørsted and Michael Faraday, the two celebrated founders of electromagnetism. Another example is Johann Ritter, the discoverer of ultraviolet radiation. Nonetheless, one cannot conclude from the case that good physics can flourish in the absence of experimental testing of theories. Neither Ørsted, Faraday, nor Ritter subscribed to Schelling's more extreme ideas and especially not to his disrespect of experiment. Ørsted's belief in a unity of electric and magnetic forces was rooted in the Romantic philosophy, but it was only when he verified it experimentally in 1820 that he turned it into a scientific discovery.

More than a century later we meet a very different version of rationalistic physics in the context of "cosmophysics," an ambitious attempt to formulate a complete and final theory of the universe and all what is in it. The leading cosmophysicists of the 1930s were two of Britain's most reputed scientists, Arthur Eddington and E. Arthur Milne. Although their world systems were quite different, they had in common that they aimed at reconstructing the very foundation of physics; they did so by basing physics on a priori principles from which the laws of nature could be uniquely deduced by pure reason. Experimental tests played but an insignificant role, being subordinated logical and mathematical arguments. Milne seriously believed that when his system of world physics (as he called it) was completed there would be no contingent elements at all in the laws of nature; it would then turn out that the laws were no more arbitrary than the theorems of geometry. A mathematician knows whether a theorem is true or not without consulting nature. Likewise, Milne (1948, p. 10) wrote that "it is sufficient that the structure [of world physics] is self-consistent and free from contradiction."

Eddington's idiosyncratic fundamental theory promised a way to deduce unambiguously all the laws and constants of nature from epistemic and mathematical considerations. In his bold attempt to unify cosmology and the quantum world, mathematics played a role no less elevated than in Milne's theory (Eddington, 1936, p. 3; Durham, 2006; Kragh, 2017):

> It should be possible to judge whether the mathematical treatment and solutions are correct, without turning up the answer in the book of nature. My task is to show that our theoretical resources are sufficient and our methods powerful enough to calculate [of nature] the constants exactly – so that the observational test will be the same kind of perfunctory verification that we apply to theorems in geometry.

Of course, neither Milne nor Eddington could afford the luxury of disregarding experiments altogether. But they argued that experiments did not reveal the true laws of nature and consequently could not be used to test the laws. Eddington famously calculated the precise values of many of the constants of nature such as the fine-structure constant, the proton-to-electron mass ratio, and the cosmological constant. When experiments failed to agree with the predicted values he arrogantly maintained that the theory was correct; any discrepancy between theory and experiment must lie with the latter.

The theories of Milne, Eddington and their few followers shared the same fate as the revolutionary Romantic natural philosophers: they were unproductive mistakes and are today relegated to the long list of grand failures in the history of science. All the same they are of some relevance in so far that aspects of the same aspirations and rationalist methods can still be found in modern physics. The most extreme version is probably the Platonic "mathematical universe hypothesis" proposed by Max Tegmark (2014), but also in the history of string theory there are examples which show at least some similarity to the earlier ideas of cosmophysics. Referring to the theory of superstrings, John Schwarz (1998) wrote, "I believe that we have found the unique mathematical structure that consistently combines quantum mechanics and general relativity. So it must almost certainly be correct." Unfortunately the prediction of supersymmetric particles remained unverified, but this did not worry Schwarz too much: "For this reason, even though I do expect supersymmetry to be found, I would not abandon this theory if supersymmetry turns out to be absent."

Without going into further detail I think one can conclude from the history of physics that fundamental theories, in order to be successful from a physical (and not merely mathematical) point of view, must have some connection to empirical reality. The historical record of such theories suggests that empirical testability is a necessary condition for progress. But this is as far as the historical argument can go. Because one can observe some regularity in the past – say that all physically progressive theories have been actually testable – there is no guarantee that the regularity will continue in the future.

Many of the arguments in string theory and multiverse physics rely implicitly on two philosophical principles which can be traced back to Leibniz in the late seventeenth century. One is the doctrine of a pre-established harmony between the mathematical and physical sciences, making pure mathematics the royal road to progress and unification in physics (Kragh, 2015). The other is the principle of plenitude which essentially states that whatever is conceived as possible must also possess physical reality. The plenitude principle is a metaphysical claim that translates potential existence into real existence. In its more modern formulation it is often taken to mean that theoretical entities exist in nature in so far that they are consistent with the fundamental laws of physics. Since numerous other universes than ours are consistent with the equations of string theory they must presumably exist (Susskind, 2006, p. 268).

The ontological plenitude principle has played a most important role in the history of science and ideas, including modern theoretical physics from Dirac's positron to Higgs's boson. Although in many cases it has been dramatically fruitful, it cannot be justified by reference to its historical record. For every example of success, there is one of failure. If the former are better known than the latter it is because history is written by the victors. In this case as in many others, history of science is ambiguous. It does not speak unequivocally in favour of either the principle of plenitude or a pre-established relationship between mathematics and physics; nor does it speak unequivocally against the doctrines.

**5. A Victorian analogy to string theory?**

Yet another case from the past deserves mention, namely the vortex theory of atoms which attracted much scientific attention during the latter part of the nineteenth century. In particular from a methodological point of view, but not of course substantially, there is more than a little similarity between this long forgotten theory and the string theory of contemporary physics. Based on a hydrodynamic theory of vortices proposed by Hermann von Helmholtz, in 1867 William Thomson (the later Lord Kelvin) argued that atoms and all atomic phenomena might be understood in terms of permanent vortex rings and filaments moving in a continuous, all-pervading medium or "fluid." The medium was generally identified with the commonly assumed world ether, meaning that the discrete atoms were reduced to particular states of motion in the continuous ether. Dualism was replaced by monism. For details about the history of the vortex theory, see Kragh (2002) which includes references to the sources of the quotations in this section.

For about two decades the research programme initiated by Thomson was vigorously cultivated by a minor army of mostly British physicists and mathematicians. Although the mathematically complex theory did not easily lead to results that could be compared to experiments, the vortex physicists stressed that it was empirically useful and, at least in principle, testable in the ordinary sense. Indeed, it was applied to a wide range of physical and chemical phenomena, including spectroscopy, the behaviour of gases, chemical bonds, the periodic system of the elements, and even gravitation. Considered a truly fundamental theory, ultimately it was expected to provide a physical explanation of the riddle of gravity. Thomson's collaborator Peter Tait went even further. "The theory of vortex-atoms must be rejected at once if it can be shown to be incapable of explaining this grand law of nature," he wrote in 1876. In spite of many attempts to derive gravity from the properties of vortex atoms, no explanation came forward. On the other hand, neither was it conclusively shown that the theory could not account for Newton's law of gravitation. After all, lack of verification is not falsification. Proponents of the theory suggested optimistically that, when the theory was developed into still more advanced versions it would eventually solve the problem.

Despite the vortex theory's connection to a variety of physical and chemical phenomena, its empirical record was far from impressive. And yet, although quantitative verification was missing there were just enough suggestive qualitative agreements to keep the theory alive as more than just a mathematical research programme. A characteristically vague defence of the vortex theory was offered by the American physicist Silas Holman: "The theory has not yet, it is true, been found capable of satisfactorily accounting for several important classes of phenomena, … but this constitutes no disproof. The theory must be judged by what it has accomplished, not by what we have not yet succeeded in doing with it."

What justified the vortex theory and made it so attractive, was not so much its ability to explain and predict physical phenomena as it were its methodological and aesthetic virtues. Albert Michelson believed that the vortex theory "ought to be true even if it is not;" and Oliver Lodge similarly described it as "a theory about which one may almost dare to say that it deserves to be true." It was also these virtues – the theory's purity and lack of arbitrary hypotheses – which greatly appealed to Maxwell. "When the vortex atom is once set in motion, all its properties are absolutely fixed and determined by the laws of motion of the primitive fluid, which are fully expressed in the fundamental equations," he wrote in 1875. "The method by which the motion of this fluid is to be traced is pure mathematical analysis. The difficulties of the method are enormous, but the glory of surmounting them would be unique."

Mathematics played a most essential role in how the vortex theory was developed and perceived. In fact, a large part of the development was driven by interest in pure mathematics and quite unrelated to physical phenomena. Papers on the subject appeared equally in journals devoted to mathematics and physics. In his 1867 paper Thomson pointed out that the calculation of the vibration frequencies of a vortex atom presented "an intensely interesting problem of pure mathematics." To his mind and to the minds of many other vortex theorists, the difficulties of deducing observable phenomena from the vortex theory were a challenge rather than an obstacle. The difficulties only added to the appeal of the theory.

The hope that progress would come through mathematics was a persistent theme in the history of the theory. It was routinely argued that it was not *yet* understood sufficiently to be physically useful. According to Tait, theoretical investigations of the kind that related to real phenomena would "employ perhaps the lifetimes for the next two or three generations of the best mathematicians in Europe." Although this was a formidable difficulty, there was no reason for despair. Because, "it is the only one which seems for the moment to attach to the development of this extremely beautiful speculation; and it is the business of mathematicians to get over difficulties of this kind."

The reliance on mathematics was a mantra among the advocates of the vortex theory of matter and ether. Here is yet another expression of the mantra, this time from the British physicist Donald Mcalister:

> The work of deduction is so difficult and intricate that it will be long before the resources of the theory are exhausted. The mathematician in working it out acquires the feeling that, although there are still some facts like gravitation and inertia to be

> explained by it, the still unexamined consequences may well include these facts and others still unknown. As Maxwell used to say, it already explains more than any other theory, and that is enough to recommend it. The vortex-theory is still in its infancy. We must give it a little time.

Although the term "theory of everything" did not exist at the time of the vortex theory (it seems to date from 1985), this is what the theory aimed to be. As late as 1895 – just a few years before the arrival of quantum theory – the mathematical physicist William Hicks discussed the vortex theory as far the best candidate for what he called an ultimate theory of pure science. The aim of such a theory, he said, was "to explain the most complicated phenomena of nature as flowing by the fewest possible laws from the simplest fundamental data." Science, he went on, "will have reached its highest goal when it shall have reduced ultimate laws to one or two, the necessity of which lies outside the sphere of our cognition." When the laws had been found, "all physical phenomena will be a branch of pure mathematics." Hicks believed that the vortex theory was at least a preliminary version of the ultimate theory, a complete version of which would require even more mathematical investigation. "It is at present a subject in which the mathematicians must lead the attack."

The mathematical richness of the vortex theory might be considered a blessing, since it effectively protected the theory from being falsified, but it was also a curse. It made G. F. FitzGerald exclaim that "it seems almost impossible but that an explanation of the properties of the universe will be found in this conception." The final vortex theory that he and others dreamt of could in principle explain everything, including the properties of the one and only universe. But could it also explain why the numerous other conceivable states of the universe, all of them consistent with the theory's framework, did not exist? Could it explain why the speed of light has the value it has rather than some other value? The theory could in principle explain the spectral lines and the atomic weight of chlorine, but had chlorine had any other spectrum and other atomic weight the theory could account for that as well. In short, the ratio between the theory's explanatory and predictive force was embarrassingly large.

Let me end with one more quotation: "I feel that we are so close with vortex theory that – in my moments of greatest optimism – I imagine that any day, the final form of the theory might drop out of the sky and land in someone's lap. But more realistically, I feel that we are now in the process of constructing a much deeper theory of anything we have had before and that … when I am too old to have any useful thoughts on the subject, younger physicists will have to decide whether we have in fact found the final theory!" The quotation is not from a Victorian physicist in the late nineteenth century but from an interview Edward Witten gave in the late twentieth century. The only change I have made is to substitute "vortex" for "string" in the first sentence. It is probably unnecessary to elaborate on the methodological and rhetorical similarities of the vortex theory of the past and the string theory of the present.

I find the case of the vortex theory to be instructive because it exhibits on a meta-level some of the features that are met in much later fundamental theories of physics. Among these are the seductive power of mathematics and the no less seductive dream of a final theory. It illustrates that some of the general problems of contemporary physics are not specifically the result of the attempts to unify quantum mechanics and general relativity. But the dissimilarities are no less distinct than the similarities. For example, string theory is cultivated on a much larger scale than the vortex theory, which largely remained a British specialty. Another difference is the ideological and religious use of the vortex theory in Victorian Britain, where it was customary to see the initial vortical motion in the ether as a result of God's hand. As Thomson pointed out in his 1867 address, vortex atoms could only have come into existence through "an act of creative power." As far as I know, the religious dimension is wholly absent from string theory (but not always from multiverse cosmology).

## 6. Conclusion

The primary function of history of science is to describe and understand past science irrespective of whether or not it connects with the modern development. It is not to assist or guide contemporary science, a function for which historians of science are generally as unfit as philosophers are. There is no reason to believe that modern string theorists would perform better if they were well acquainted with earlier episodes in the history of physics. While this is undoubtedly true in a technical sense, historical reflection has something to offer in a broader sense, namely when it comes to the general understanding of the present debate about fundamental physics. In my view, what can be learned from the past relates mostly to the philosophical aspects of the debate, to the novelty of the epistemic situation, and to the rhetoric used in the presentation of modern fundamental physics to a general public.

By adopting a historical perspective it becomes clear that many of the claims concerning the uniqueness of the present situation are exaggerated. It is not the first time in history that scientists have seriously questioned the traditional norms of science and proposed revisions of what should pass as theory assessment. Apart from the precedents mentioned here, there are several other cases of intended epistemic shifts (Kragh, 2011). Physics may presently be at "a turning point" or "a remarkable paradigm shift," but if so it is not for the first time. Knowledge of the history of science and the history of fundamental physics in particular, cannot but result in an improved and more balanced picture of what is currently taking place in areas such as string theory and multiverse cosmology.

**References**

Barrau, A. (2007). Physics in the multiverse. *Cern Courier* (20 October).

Burgess, C. & Quevedo, F. (2007). The great cosmic roller-coaster ride. *Scientific American* **297** (10), 29-35.

Cunningham, A. & Jardine, N., eds. (1990). *Romanticism and the Sciences*, Cambridge: Cambridge University Press.

Dawid, R. (2013). *String Theory and the Scientific Method*, Cambridge: Cambridge University Press.

Donovan, A., Laudan L. & Laudan, R., eds. (1988). *Scrutinizing Science: Empirical Studies of Scientific Change*, Dordrecht: Kluwer.

Durham, I. (2006). Sir Arthur Eddington and the foundation of modern physics. Arxiv:quant-ph/0603146.

Eddington, A. (1936). *Relativity Theory of Protons and Electrons*, Cambridge: Cambridge University Press.

Einstein, A. (1950). On the generalized theory of relativity. *Scientific American* **182** (4), 13-17.

Ellis, G. & Silk, J. (2014). Defend the integrity of physics. *Nature* **516**, 321-3.

Gold, T. (1956). Cosmology. *Vistas in Astronomy* **2**, 1721-26.

Johannson, L. G. & Matsubaru, K. (2009). String theory and general methodology: A mutual evaluation. *Studies in History and Philosophy of Modern Physics* **42**, 199-210.

Kragh, H. (2002). The vortex atom: A Victorian theory of everything. *Centaurus* **44**, 32-114.

Kragh, H. (2011). *Higher Speculations: Grand Theories and Failed Revolutions in Physics and Cosmology*, Oxford: Oxford University Press.

Kragh, H. (2015). Mathematics and physics: The idea of a pre-established harmony. *Science and Education* **24**, 515-27.

Kragh, H. (2017). Eddington's dream: A failed theory of everything. In I. Durham & D. Rickles, eds., *Information and Interaction*, Basel: Springer.

Kuhn, T. (1977). *The Essential Tension: Selected Studies in Scientific Tradition and Change*, Chicago: University of Chicago Press.

Laudan, L. (1977). *Progress and Its Problems: Towards a Theory of Scientific Growth*, London: Routledge & Kegan Paul.

Matthews, R. (2008). Some swans are grey. *New Scientist* **198**, 44-7.

Milne, E. A. (1948). *Kinematic Relativity*, Oxford: Clarendon Press.

Schellekens, A. N. (2008). The emperor's last clothes? Overlooking the string theory landscape. *Reports on Progress in Physics* **71**, 072201.

Schickore, Jutta (2011). More thoughts on HPS: Another twenty years later. *Perspectives on Science* **19**, 453-81.

Schwarz, J. H. (1998). Beyond gauge theories. Arxiv:hep-th/9807195.

Shapere, D. (2000). Testability and empiricism. In E. Agazzi & M. Pauri, eds., *The Reality of the Unobservable*, Dordrecht: Kluwer Academic.

Smolin, L. (2004). Atoms of space and time. *Scientific American* **290** (1), 66-75.

Susskind (2006). *The Cosmic Landscape: String Theory and the Illusion of Intelligent Design*, New York: Little, Brown and Company.

Tegmark, M. (2007). Many lives in many worlds. *Nature* **448**, 23-4.

Tegmark, M. (2014). *Our Mathematical Universe*, New York: Alfred E. Knopf.

Weinberg, S. (2007). Living in the multiverse. In B. Carr, ed., *Universe or Multiverse?* Cambridge: Cambridge University Press.

Yang, A. (2008). Matters of demarcation: Philosophy, biology, and the evolving fraternity between disciplines. *International Studies in the Philosophy of Science* **22**, 211-25.